\documentstyle[amscd,amssymb,verbatim]{amsart}
\begin{document}
\numberwithin{equation}{section} \title{A Modification of Nambu's
  Mechanics} 
\author{Peter Varga}
\address{Institute of Mathematics and Informatics \\
  Lajos Kossuth University\\
  H-4010 Debrecen, Hungary} 
\email{varga@@math.klte.hu} \thanks{}
\date{}

\begin{abstract}
  The Poisson, contact and Nambu brackets define algebraic structures
  on $C^{\infty}(M)$ satisfying the Jacobi identity or its
  generalization.  The automorphism groups of these brackets are the
  symplectic, contact and
  volume preserving diffeomorphism groups. We introduce a modification
  of the Nambu bracket, which defines an evolution equation generating
  the whole diffeomorphism group.  The relation between the modified
  and usual Nambu brackets is similar to the relation between the
  Poisson and contact structures. We briefly discuss the problem of
  quantization of the modified bracket.
\end{abstract}

\maketitle

\section{Introduction} \label{S:intro}
In 1973 Nambu constructed a generalization of Hamiltonian mechanics
\cite{N}. He defined a dynamical system on ${\Bbb{R}}^{3}$ by the
trilinear Nambu bracket:
\begin{equation}\label{E:i1}
\frac{df}{dt}=\{ {H_{1},H_{2},f} \}=
\frac{\partial{\left({H_{1},H_{2},f}\right)}}
     {\partial{\left({x,y,z}\right)}}=
L_{H_{1},H_{2}}f,
\end{equation}
where the third term is the Jacobian of $(H_{1},H_{2},f)$ with respect
to $(x,y,z)$, and $L_{H_{1},{H_{2}}}$ is a vector field on ${\Bbb{R}}^{3}$. The
flow $\phi_{t}$ generated by $L_{H_{1},{H_{2}}}$ is a canonical
  transformation in the sense that
\begin{equation}\label{E:i2}
\{H_{1}\circ\phi_{t},H_{2}\circ\phi_{t},f\circ\phi_{t}\}=
\{{H_{1},H_{2},f}\}\circ\phi_{t}
\end{equation}
is satisfied.
Since the Nambu bracket can be defined in terms of the volume form of
${\Bbb{R}}^{3}$ \eqref{E:i2} holds for any volume preserving transformation
$\phi_{t}$, and indeed $\operatorname{div}{L_{H_{1},{H_2}}}=0$. 
 So Nambu dynamics
is related to volume forms and volume preserving transformations just
as Hamiltonian dynamics relates to symplectic two-forms $\omega$ and
symplectic transformations leaving $\omega$ invariant.

Our main result is the introduction of a new bracket which generates
transformation $\phi_{t}$ satisfying \eqref{E:i2} although $\phi_{t}$
does not leave the volume form invariant. In some sense, the new
bracket generates the whole diffeomorphism group.  At the first sight
the existence of such bracket seems to be unlikely, since no tensor is
invariant under the action of the diffeomorphism group. However, the
existence of the group of contact transformation and the contact
bracket shows that strict invariance might be unnecessary, as contact
transformations carry the contact one-form $\alpha$ into its scalar
multiple $f\alpha$. The volume form $\nu$ behaves the same way under
the action of a general diffeomorphism $\phi_{t}$:
$\phi^{*}_{t}\nu=f\nu$. As the space of top dimensional differential
forms at a point is one dimensional, the existence of $f$ is
completely trivial!

We construct a trilinear bracket 
\begin{equation} 
[\,,\,,\,]:\,C^{\infty}(R^{2})\otimes{}C^{\infty}(R^{2})\otimes{}
C^{\infty}(R^{2})\longrightarrow{}
C^{\infty}(R^{2})
\end{equation}
(called the
modified Nambu bracket) which fulfills \eqref{E:i2}:
\begin{equation}\label{E:i3}
[H_{1},H_{2},f]_{mN}=H_{1}\{H_{2},f\}-H_{2}\{H_{1},f\}+f\{H_{1},H_{2}\}.  
\end{equation}  
($\{,\}$is the standard Poisson bracket on ${\Bbb{R}}^{2}$.) The
 vector fields
$L_{H_{1},H_{2}}$ defined by
\begin{equation}\label{E:i4} 
L_{H_{1},H_{2}}f=H_{1}\{H_{2},f\}-H_{2}\{H_{1},f\}
\end{equation}
generates the whole diffeomorphism group. Just as the Nambu bracket
\eqref{E:i1}, this construction can generalized to ${\Bbb{R}}^{n}$.

To put our work into perspective, in Section 2 we briefly review the
properties of the Poisson, contact and Nambu brackets. In Section 3
we study the modified Nambu bracket. Section 4 present an embedding
of the modified Nambu dynamics into Nambu's original one. Section 5
contains some remarks on the problem of quantization of the new
bracket.

\section{Brackets on $C^{\infty}(M)$.}\label{S:brackets}

In this section we review the properties of the Poisson, contact and
Nambu brackets.

The most well-known bracket is the Poisson bracket on symplectic
manifolds $(M^{2n},\omega)$ where $\omega$ is the symplectic two-form:
\begin{equation}\label{E:b1a}
\{f,g\}_{P}=\eta(df,dg).
\end{equation}
The two-vector $\eta$ is the inverse of the map $\omega : TM \to
T^{*}M$, defined as $\omega(X)=i_{X}\omega$.  The Poisson bracket
satisfies
\begin{equation}\label{E:b2a}
\{f,g\}_{P} = -\{g,f\}_{P}, 
\end{equation}
\begin{equation}\label{E:b2b}
\{f,\{g,h\}\}_{P} = \{\{f,g\},h\}_{P}+\{g,\{f,h\}\}_{P}, 
\end{equation}
\begin{equation}\label{E:b2c}
\{f,gh\}_{P} = \{f,g\}h+\{f,h\}g.  
\end{equation}
Leibniz's rule \eqref{E:b2c} implies that $\{f,g\}=X_{f}g$ ($X_{f}\in
TM$).  As a consequence of the Jacobi identity \eqref{E:b2b}
$\phi_{t}^{H}=\exp{tX_{H}}$ satisfies
\begin{equation} \label{E:b3a}
\{f \circ \phi_{t}^{H}, g\circ \phi_{t}^{H}\}=\{f,g\}\circ \phi_{t}^{H}.
\end{equation}
\eqref{E:b3a} holds since $\omega$ (and $\eta$) is invariant under the
flow $ \phi_{t}^{H}$.

The next case is the contact bracket \cite{Hu} . It is defined
on contact manifolds $(M^{2n+1},\alpha)$ where $\alpha$ is the contact
one form ($ \alpha \in \Lambda^{1}M, \alpha \wedge
(d\alpha)^{n}\ne0$). The contact one-form is defined only up to a
scalar factor, so the contact structure is more invariantly defined by
the field of $2n$ dimensional planes annihilating $\alpha$.  The
infinite dimensional contact transformation group leaves the plane
field invariant, and carries $\alpha$ into $ f\alpha$ for some 
$f\in{}C^{\infty{}}(M^{2n+1})$. 
By a suitable
coordinate transformation $\alpha$ can be put into its normal form
\begin{equation} \label{E:b4}
\alpha=dx^{2n+1}+\sum_{i=1}^{n}{x^{i}dx^{i+n}}.
\end{equation}
In this coordinate system the contact bracket is
\begin{align} \label{E:b5}
  \{f,g\}_{cont}= \sum_{i=1}^{n}
  \bigl(\partial_{i}{f}\partial_{i+n}{g}-
    \partial_{i}{g}\partial_{i+n}{f}\bigr) &+
  \biggl(\partial_{2n+1}{f}\sum_{i=1}^{2n}{x_{i}\partial_{i}{g}-2g}\biggr)\\
 &-  \biggl(\partial_{2n+1}{g}\sum_{i=1}^{2n}{x_{i}\partial_{i}{f}-2f}\biggr).
\end{align}
\eqref{E:b5} satisfies the Jacobi identity but not Leibniz's   rule
since \eqref{E:b5} has the structure
\begin{equation} \label{E:b6}
\{f,g\}_{cont}=K(f)g+c(f)g=H(f)g,
\end{equation}
where $K(f)\in{TM}$ and $c(f)\in{C^{\infty}(M)}$, and the appearance
of the multiplier term $c(f)$ is incompatible with Leibniz's rule.

The theory of Jacobi manifolds unifies the concepts of symplectic and
contact manifolds \cite{K}. The Jacobi bracket is given by
\begin{equation} \label{E:b7}
\{f,g\}_{J}=\eta(df,dg)+fEg-gEf,
\end{equation}
where the $\eta\in{}\Lambda{}^{2}TM$ bivector
 and the $E\in{}TM$ vector field satisfy 
compatibility conditions necessary for the fulfillment of the Jacobi 
 identity
\begin{equation} \label{E:b8}
[\eta,\eta]=2E\wedge\eta, [E,\eta]=L_{E}\eta=0,
\end{equation}
where $[\,,]$ is the Schouten-Nijenhuis bracket \cite{V}.  
Jacobi manifolds are locally decomposable into an union of symplectic and
contact leaves.

A more recently introduced bracket operation is the generalized Nambu
bracket on $R^{n}$:
\begin{equation}\label{E:n1}
\begin{aligned}
&\{\,,\dots{},\,\}:\,C^{\infty{}}(T{\Bbb{R}}^{n})^{\otimes{}n}
\longrightarrow{}
C^{\infty{}}(T{\Bbb{R}}^{n})\\
&\{f_{1},f_{2},\dots,f_{n}\}_{N}=
\epsilon_{i_{1}\dots i_{n}}\partial_{i_{1}}{f_{1}}\dots 
\partial_{i_{1}}{f_{n}}=\eta_{0}^{(n)}(df_{1},\dots,df_{n}),
\end{aligned}
\end{equation}
where $\epsilon{}$ is the alternating Levi-Civita symbol.
The Nambu bracket satisfies
\begin{equation}\label{E:n2a}
\{f_{1},\dots,f_{i},\dots,f_{j},\dots,f_{n}\}_{N}=
-\{f_{1},\dots,f_{j},\dots,f_{i},\dots,f_{n}\}_{N},
\end{equation}
\begin{equation}\label{E:n2b}
\{H_{1},\dots,H_{n-1},gh\}_{N}=h\{H_{1},\dots,H_{n-1},g\}_{N}+
g\{H_{1},\dots,H_{n-1},h\}_{N},
\end{equation}
\begin{equation}\label{E:n2c}
\{H_{1},\dots,H_{n-1},\{g_{1},\dots,g_{n}\}_{N}\}_{N}=
\sum_{i=1}^{n}{  \{g_{1},\dots,\{H_{1},\dots,H_{n-1},g_{i}\}_{N},
\dots,g_{n}\}_{N}}.
\end{equation}
At this point we remark that different generalizations of the Jacobi
identity gained some popularity recently \cite{hanlon,az1,az2,az3,perez}.
Just as for the Poisson-bracket, \eqref{E:n2b} ensures that
\begin{equation}\label{E:n3}
\{H_{1},\dots,H_{n-1},g\}_{N}=X_{H_{1},\dots,H_{n-1}}g,
\end{equation}
for some vector field $X_{H_{1},\dots,H_{n}} \in{T{\Bbb{R}}^{n}}$.
 Since the Fundamental
Identity  (FI) \eqref{E:n2c} holds \cite{T,DFST},
$\phi_{t}=\exp{tX_{H_{1},\dots,H_{n}}}$ is a canonical transformation
(an automorphism of the bracket):
\begin{equation}\label{E:n4}
\{f_{1},\dots,f_{n}\}_{N}\circ\phi_{t}{}=
\{f_{1}\circ{}\phi_{t}{},\dots,f_{n}\circ{}\phi_{t}{}\}_{N}.
\end{equation}
Indeed, \eqref{E:n2c} is the derivative of \eqref{E:n4}. As
\begin{equation}\label{E:n5}
\mathop{div}X_{H_{1},\dots,H_{n-1}}=
\partial{}_{i_{n}}\bigl(
\epsilon{}_{i_{1}\dots{}i_{n}}\partial{}_{i_{1}}H_{1}
\dots{}\partial{}_{i_{n-1}}H_{n-1}
\bigr){}=0,
\end{equation}
$\phi_{t}$ leaves the standard volume form
$\nu_{0}{}=dx^{1}\wedge\dots\wedge{}dx^{n}$ invariant. 

Nambu dynamics
can be formulated in terms of the volume form $\nu{}$ instead of the
n-vector $\eta{}$.  For that purpose, let us define $\nu_{\eta{}}$ by
the condition $\nu_{\eta}{}(\eta{})=1$. For a given set of Nambu 
Hamiltonians
$H_{1},\dots{},H_{n-1}$ the vector field $X_{H_{1},\dots,H_{n-1}}$ can
be obtained by the following process:       First find $n-1$ vectors
$X_{1},\dots{},X_{n-1}$ satisfying
$\langle{}X_{i},dH_{j}\rangle{}=\delta_{i,j}{}$ . Then
$X_{H_{1},\dots,H_{n-1}}$ is determined by the conditions
\begin{equation}\label{E:n6}
X_{H_{1},\dots,H_{n-1}}H_{i}=0,\quad\text{and}
\quad{}
\nu_{\eta}{}(X_{1},\dots{},X_{n-1},
X_{H_{1},\dots,H_{n-1}})=1.
\end{equation}
Since the Nambu bracket is determined by the volume form \eqref{E:n5}
implies \eqref{E:n4} as divergenceless vector fields leave the volume
form unchanged.  As on a manifold two volume forms are equivalent if
their total masses are equal, on ${\Bbb{R}}^{n}$ any
 nondegenerate $n$-vector  has
the form $\eta_{0}^{(n)}{}$ in a suitable coordinate system.

The vector fields $X_{H_{1},\dots,H_{n-1}}$ form only a subset of the
Lie-algebra of volume preserving transformations.  Since
\begin{equation}
\frac{d}{dt}(H_{i}\circ{}\phi{}_{t})=0,
\end{equation}
 the orbits of Nambu dynamics are
one-dimensional intersections of level-surfaces, so if the motion is
confined to a bounded region, then it must be periodic.  The set
$\{X_{H_{1},\dots,H_{n-1}}\}$ is not even a linear space, as in
general there are no $F_{1},\dots{},F_{n-1}$ fulfilling
$X_{G_{1},\dots,G_{n-1}}+X_{H_{1},\dots,H_{n-1}}=X_{F_{1},\dots,F_{n-1}}$
for a given set of $\{G_{i},H_{i}\}$ functions.Nevertheless \eqref{E:n4}
holds for any volume preserving transformations $\phi{}_{t}^{volpres}$, 
not just for
$\phi_{t}{}=\exp{X_{H_{1},\dots,H_{n-1}}}$.

\section{The modified Nambu bracket}\label{S:mod}

The General form of the Jacobi bracket \eqref{E:b7} suggests the
following generalization of the Nambu bracket:
\begin{equation}\label{E:m1}
[f_{1},\dots{},f_{n}]_{mN}=\eta{}(df_{1},\dots{},df_{n})+
\sum_{i=1}^{n}{(-1)^{i+1}f_{i}e(
df_{1},\dots{},\hat{f_{i}},\dots{},f_{n})},
\end{equation}
where $\eta{}$ and $e$ are $n$ and $n-1$ vectors. This bracket is
antisymmetric, but does not satisfies the Leibniz identity. The
Fundamental Identity imposes a set of complicated consistency
conditions on $\eta{}$ and $e$.
We prove that the following bracket on $C^{\infty}(R^{n})$satisfies
the FI:
\begin{equation}\label{E:m2}
\begin{aligned}
&[\,,\dots{},\,]:\,C^{\infty{}}(T{\Bbb{R}}^{n})^{\otimes{}(n+1)}
\longrightarrow{}
C^{\infty{}}(T{\Bbb{R}}^{n})\\
&[f_{1},\dots{},f_{n+1}]_{mN}=
\sum_{i=1}^{n+1}{(-1)^{i+1}f_{i}\eta_{0}^{(n)}{}(
df_{1},\dots{},\overset{\hat{}}{f_{i}},\dots{},f_{n+1})}.
\end{aligned}
\end{equation}
($\eta_{0}^{(n)}{}$ is the standard Nambu tensor on $R^{n}$.) 

 For the
sake of clarity, we present a proof first for the $n=2$ case.
Let $\{f,g\}_{P}=\eta_{0}^{(2)}{}(df,dg)$ the standard Poisson-bracket
on ${\Bbb{R}}^{2}$, and define the trilinear alternating modified Nambu
bracket as
\begin{align}\label{E:m3}
  [f,g,h]_{mN} &= f\{g,h\}_{P}-g\{f,h\}_{P}+h\{f,g\}_{P}\\
 &= f\{g,h\}_{P}+g\{h,f\}_{P}+h\{f,g\}_{P}.
\end{align}
Our goal is to prove
\begin{multline}\label{E:m4}
[H_{1},H_{2},[f,g,h]_{mN}]_{mN}=
[[H_{1},H_{2},f]_{mN},g,h]_{mN}\\
[f,[H_{1},H_{2},g]_{mN},h]_{mN}+[g,h,[H_{1},H_{2},f]_{mN}]_{mN}
\end{multline}
We introduce the shorthand notation
\begin{equation}\label{E:m5}
[H_{1},H_{2},\phi{}]_{mN}=T\phi{}=L\phi{}+H\phi{}.
\end{equation}
In this expression $L=X_{H_{1},H_{2}}=H_{1}X_{H_{2}}-H_{2}X_{H_{1}}$
and $H=\{H_{1},H_{2}\}_{P}$, where $X_{H_{i}}$ is the Hamiltonian vector
field generated by $H_{i}$ with respect to the Poisson-bracket. The
divergence of $L$ is
\begin{equation}\label{E:m6}
\operatorname{div}L=-2H,  
\end{equation}
so
\begin{equation}\label{E:m7}
{\cal{L}}_{L}(dx \wedge{} dy)=-2H(dx \wedge{} dy),
\bigskip{}\text{and}\bigskip{}\bigskip{}
{\cal{L}}_{L}\eta{}_{(0)}^{2}=2H\eta{}_{(0)}^{2}.
\end{equation}
We remark that the operator
 $U_{t}=\exp{tT_{H_{1},H_{2}}}$ preserve the 2-norm of 
$L^{2}({\Bbb{R}}^{2})$, since
\begin{equation}\label{E:m8}
\frac{d}{dt}\int{}h^{2}\,dx\wedge{}dy=
\int{h^{2} {\cal{L}}_{L}(dx \wedge{} dy)+(Hh)2h(dx \wedge{} dy)  }=0.
\end{equation}
To prove the Fundamental Identity we evaluate the left-hand (LHS) and 
right-hand (RHS) sides of \eqref{E:m4}.
\begin{equation}\label{E:m9}
  \bigl({}     {\cal{L}}_{L}+H\bigr){}
  \bigl({} f\eta{}_{0}^{(2)} (dg,dh) \bigr){}  +(\text{cycl.perms.})
  \tag{$\text{LHS}_{1}$}
\end{equation}
\begin{align}\label{E:m10}
  \bigl[{}\bigl({}{\cal{L}}_{L}+H\bigr){} &f \bigr]{}
  \eta{}_{0}^{(2)} \bigl({}dg,dh\bigr){}+
  f \eta{}_{0}^{(2)} \bigl({} d \bigl({}{\cal{L}}_{L}+H\bigr){}g,dh 
  \bigr){}+  \notag{}        \\
  &f \eta{}_{0}^{(2)} \bigl({} dg,d \bigl({}{\cal{L}}_{L}+H\bigr){}h \bigr){}+
  \text{(cycl.perms.)} \tag{$\text{RHS}_{1}$}
\end{align}
where  {\em{cycl.perms.}} refers to the cyclical permutations of 
$f,g$ and $h$.
The terms containing ${\cal{L}}_{L}f,\dots{}$, and 
$Hf\eta{}_{0}^{(2)}(dg,dh),\dots{}$ occurs the same way on both sides.
After the deletion of these terms the following expressions remain:
\begin{equation}\label{E:m11}
  f \bigl({} {\cal{L}}_{L}  \eta{}_{0}^{(2)}  \bigr){}
  \bigl({} dg,dh \bigr){} + \text{(cycl.perms.)} \tag{$\text{LHS}_{2}$}
\end{equation}
\begin{equation}
  f \eta{}_{0}^{(2)}\bigl(d(Hg),dh\bigr)+ 
  f \eta{}_{0}^{(2)}\bigl(dg,d(Hh)\bigr)
  +\text{(cycl.perms)}  \tag{$\text{RHS}_{2}$}
\end{equation}
These expressions are equal to
\begin{equation}
  2Hf  \eta{}_{0}^{(2)} (dg,dh) +  \text{(cycl.perms)}  \tag{$\text{LHS}_{3}$}
\end{equation}
\begin{align}
  2Hf  \eta{}_{0}^{(2)} (dg,dh) +  gf\eta{}_{0}^{(2)} (dH,dh) + 
  hf\eta{}_{0}^{(2)} (dg,dH) + \text{(cycl.perms)}  \tag{$\text{RHS}_{3}$}
\end{align}
However, the extra two terms of the RHS drops out. For example,
$hf\eta{}_{0}^{(2)} (dg,dH)$ is annihilated by $fh\eta{}_{0}^{(2)} (dH,dg)$
which is generated by $ h\eta{}_{0}^{(2)} (d(Hf),dg)$.
So  $\text{LHS}_{3}=\text{RHS}_{3}$, consequently LHS=RHS, i.e.  
the Fundamental Identity has been proven.

The same line of reasoning works for the bracket on ${\Bbb{R}}^{n}$ ,too.
To prove
\begin{equation}\label{E:m12}
  [H_{1},\dots{},H_{n},[f_{1},\dots{},f_{n+1}]_{mN}]_{mN}=
  \sum_{i=1}^{n} 
  [f_{1},\dots{},[H_{1},\dots{},H_{n},f_{i}]_{mN},\dots{},f_{n+1}]_{mN}
\end{equation}
we introduce
\begin{equation}\label{E:m13}
  T=T_{ H_{1},\dots{},H_{n} }=
  \sum_{i=1}^{n} (-1)^{i+1} H_{i} X_{ H_{1},\dots{},\hat{H_{i}},\dots{},H_{n} }
  + \{H_{1},\dots{},H_{n}\}_{N}=L+H, 
\end{equation}
where $X_{ H_{1},\dots{},H_{n} }$ is the vector field defined by the
condition $X_{ H_{1},\dots{},H_{n} }f=[H_{1},\dots{},H_{n},f]_{N}$.

The proof of \ref{E:m12} is very similar to the previous case.
The difference between the left-hand side (LHS) and the right-hand
side (RHS) of  \ref{E:m12} evaulates to $0$:
\begin{equation}\notag{}
  \begin{aligned}
    &\bigl\{{}\text{LHS}\bigr\}{}-
    \bigl\{{}\text{RHS}\bigr\}{}\\
    &=
    \Biggl\{{}
    \bigl({}     {\cal{L}}_{L}+H\bigr){}
    \biggl[{}
    \sum_{i=1}^{n+1}(-1)^{i+1}f_{i}
    \eta{}_{0}^{(n)}(df_{1},\dots{},\hat{df_{i}},\dots{},df_{n+1})
    \biggr]{}
    \Biggr\}{}\\
    &-
    \Biggl\{{}
    \sum_{i=1}^{n+1}(-1)^{i+1}
    \biggl[{}
    \bigl({}     {\cal{L}}_{L}+H\bigr){}f_{i}
    \biggr]{}
    \eta{}_{0}^{(n)}(df_{1},\dots{},\hat{df_{i}},\dots{},df_{n+1})\\
    &\phantom{=}+
    \sum_{i=1}^{n+1}(-1)^{i+1}f_{i}
    \sum_{k=1,k\neq{}i}^{n+1}\eta{}_{0}^{(n)}
    (df_{1},\dots{}, d( {\cal{L}}_{L}+H)f_{k},\dots{},     
    \hat{df_{i}},\dots{},df_{n+1})
    \Biggr\}{}\\
    &=
    \Biggl\{{}
    \sum_{i=1}^{n+1}(-1)^{i+1}f_{i}
    \bigl({} {\cal{L}}_{L}\eta{}_{0}^{(n)}   \bigr){}
    (df_{1},\dots{},\hat{df_{i}},\dots{},df_{n+1})
    \Biggr\}{}\\
    &-
    \Biggl\{{}
    \sum_{i=1}^{n+1}(-1)^{i+1}f_{i}
    \sum_{k=1,k\neq{}i}^{n+1}\eta{}_{0}^{(n)}
    (df_{1},\dots{}, d(Hf_{k}),\dots{} ,    
    \hat{df_{i}},\dots{},df_{n+1})
    \Biggr\}{}\\
    &=
    \Biggl\{{}
    \sum_{i=1}^{n+1}(-1)^{i+1}f_{i}
    (nH\eta{}_{0}^{(n)})
    (df_{1},\dots{},\hat{df_{i}},\dots{},df_{n+1})
    \Biggr\}{}\\
    &-
    \Biggl\{{}
    \sum_{i=1}^{n+1}(-1)^{i+1}f_{i}
    \sum_{k=1,k\neq{}i}^{n+1}
    H\eta{}_{0}^{(n)}
    (df_{1},\dots{},df_{k},\dots{},\hat{df_{i}},\dots{},df_{n+1})
    \Biggr\}{}\\
    &=0.
  \end{aligned}
\end{equation}

The FI means that $T=L+H$ is a derivation (infinitesimal automorphism) of the 
bracket. The  proof of the FI used the fact that $T$ is generated by
the Hamiltonians $H_{1},\dots{},H_{n}$ only for the computation of
$\operatorname{div}L$. Consequently any $T=L+H$ is a generator of an 
automorphism of the bracket if $\operatorname{div}L=-nH$ holds.
So the assignment $V\to{}T_{V}=V-1/n\operatorname{div}V$ maps any
 vector field to an infinitesimal automorphism  of the modified Nambu bracket.

\section{Embedding into Nambu's dynamics}\label{S:bed}

The Nambu bracket is a special case of the modified bracket:
\begin{equation}\label{E:e1}
\{f_{1},f_{2},\dots{},f_{n}\}=[1,f_{1},f_{2},\dots{},f_{n}]_{mN}.
\end{equation}
In this section we study the reverse relation and express the
 modified bracket on ${\Bbb{R}}^{n}$ with the help of the Nambu bracket on
${\Bbb{R}}^{n+1}$.

The evolution equation of the modified Nambu dynamics 
\begin{equation}\label{E:e2}
\frac{d}{dt}f=[H_{1},\dots{},H_{n},f]_{mN}=(L+H)f
\end{equation}
is not generated just by the vector field $L$ but contains a scalar term
$H$, too. Nevertheless, by adding an extra dimension to the phase space,
the evolution of $f$ can be described with a flow on the 
 extended phase space. On the extended coordinate system 
$(x_{1},\dots{},x_{2},l)$ the induced generator of the evolution is
$T^{(l)}=L+Hl\partial{}_{l}$. The flow of $T^{(l)}$ deforms the surface
$l=f(x_{1},\dots{},x_{n})$ just as the graph 
$\{(x_{1},\dots{},x_{n},f(x_{1},\dots{},x_{n}))\}$
of $f$ changes. Unfortunately,
$\operatorname{div}T^{(l)} \neq{} 0$, so $T^{(l)}$ is not a vector
field
generated by some Hamiltonians. However, if a new coordinate system is
introduced, where 
 $(x_{1},\dots{},x_{2},l)$
is mapped to
$(x_{1},\dots{},x_{2},k),\,\,k=l^{n}$ then $T^{(l)}$ is mapped to
$T^{(k)}=L+nHk\partial{}_{k}$, which is divergenceless with respect to
the volume element $dx_{1}\wedge{}\dots{}\wedge{}dx_{n}\wedge{}dk$.
$T^{(k)}$ is generated by the Nambu Hamiltonians
\begin{equation}\label{E:e3}
h_{1}=H_{1}\sqrt[n]{nk},\dots{},h_{n}=H_{1}\sqrt[n]{nk},
\end{equation}
as these Hamiltonians generate terms like
\begin{align}\label{E:e4}
&\partial{}_{{x}_{i_{1}}}\bigl({}H_{1}\sqrt[n]{nk}\bigr){}\dots{}
\partial{}_{{x}_{i_{n}}}\bigl({}H_{n}\sqrt[n]{nk}\bigr)\partial{}_{k},
 \notag{}\\
&\partial{}_{{x}_{i_{1}}}\bigl({}H_{1}\sqrt[n]{nk}\bigr)\dots{}
 \partial{}_{k}\bigl({}H_{j}\sqrt[n]{nk}\bigr)\dots{}
 \partial{}_{x_{i_{j}}}
\propto{}
H_{j}\partial{}_{{x}_{i_{1}}}H_{1}\dots{}
 \partial{}_{x_{i_{j}}}
\end{align}
The volume form 
$\nu{}^{(k)}=dx_{1}\wedge{}\dots{}\wedge{}dx_{n}\wedge{}dk$
 and the functions $h_{i}$
can be pulled back to the $(x_{1},\dots,x_{n},l)$ coordinate system:
\begin{equation}\label{E:e5}
\nu{}^{(l)}=nl^{n-1}dx_{1}\wedge{}\dots{}\wedge{}dx_{n}\wedge{}dl,
\bigskip{}\bigskip{} h_{i}^{(l)}=\sqrt[n]{n}lH_{i}.
\end{equation}
The Nambu $(n+1)$-vector corresponding to $\nu{}^{(l)}$ is
\begin{equation}
\eta{}^{(l)}=\frac{1}{n}l^{1-n}
\partial{}_{x_{1}}\wedge{}\dots{}\partial{}_{x_{n}}\wedge{}\partial{}_{l}.
\end{equation}
$\eta{}^{(l)}$ and $h_{i}^{(l)},\,i=1..n$ generate the desired flow
$T^{(l)}$ by the Nambu bracket on ${\Bbb{R}}^{n+1}$.

The presented one dimensional extension of the phase space is very similar 
 to the symplectification of contact manifolds. We re-derive the previous
 result using the framework of \cite{A}.  Let $(M,\eta{})$  be a Nambu 
manifold with $n$-vector $\eta{}$. Consider the line bundle
 $\pi{}:L \to{}M$ where the fiber over a point $x\in{}M$ is the set
 $\{ \lambda{}\eta{}(x),\lambda{}\in{}{\Bbb{R}}^{+}\}$. $L$ is a
 subbundle of $\Lambda{}^{n}TM$. (If $\eta{}$ vanish at some points, 
we can discard those points as there the value of the Nambu bracket is
 zero anyway.) Define the function $y \in{}C^{\infty{}}(L)$ by the conditions 
$y(x_{i},\lambda{}\rho{})=\lambda{}y(x_{i},\rho{}),\quad{} y(x_{i},\eta{})=1$.
  The one-homogeneous extension of $f$ is 
 $\tilde{f}=y\pi{}^{\ast{}}(f)$.
We introduce on $L$ an $(n+1)$-vector Nambu tensor $\tilde{\eta{}}$ by the 
 condition
\begin{equation}\label{E:c1}
\tilde{\eta{}}(dy,d\tilde{f_{1}},\dots{},d\tilde{f_{n}})=
\widetilde{\eta{}(df_{1},\dots{},df_{n})}.
\end{equation}
To prove the existence of $\tilde{\eta{}}$, we note that
\begin{equation}\label{E:c2}
\tilde{\eta{}}(dy,d\tilde{f_{1}},\dots{},d\tilde{f_{n}})=
 y^{n}\tilde{\eta{}}(dy,\pi{}^{\ast{}}(f_{1}),\dots{},
\pi{}^{\ast}{}(f_{n})),
\end{equation}
so the following tensor satisfies \eqref{E:c1}:
\begin{equation}\label{E:c3}
\tilde{\eta{}}=\frac{1}{y^{n-1}}\partial{}_{vert}\wedge{}\eta{}_{ext},
\end{equation}
where $\eta{}_{ext}$ is any $n$-vector on $L$ satisfying
\begin{equation}\label{E:c4}
\eta{}_{ext}(d\pi{}^{\ast{}}(f_{1}),\dots{},d\pi{}^{\ast{}}(f_{n}))=
\eta{}(df_{1},\dots{},df_{n}),
\end{equation}
and $\partial{}_{vert}$ is a vertical vector  such that 
 $\partial{}_{vert}y=1$ 
   The 'pull-back' $\eta{}_{ext}$ of $\eta{}$ is determined only up to
 terms containing $\partial{}_{vert}$, but these terms drop out of
 \eqref{E:c3} anyway. Now we define the modified bracket by
\begin{equation}\label{E:c5}
[f_{1},\dots{},f_{n+1}]_{mN}(x_{i})=
\tilde{\eta{}}(d\tilde{f_{1}},\dots{}d\tilde{f}_{n+1})(x_{i},\eta{}).
\end{equation}
Since $\tilde{\eta{}}(d\tilde{f_{1}},\dots{}d\tilde{f_{n+1}})$ is 
one-homogeneous, the modified brackets inherits the Fundamental Identity
 from the Nambu bracket of $L$. (Leibniz's rule is not inherited, since 
it contains the product of two functions ruining one-homogeneity).

These constructions provide a second proof of the FI since we mapped 
the modified bracket onto the Nambu bracket which satisfies the FI.
It show the possibility of an action formulation for the modified Nambu 
dynamics \cite{T}.

\section{On the quantization of the modified Nambu bracket}\label{S:quant}

The quantization of the Nambu bracket is not a terribly well-defined task,
let alone the modified bracket. In this section we make a few remarks on
 the connection between the quantization of the two brackets. Since the
 modified and
the usual Nambu mechanics can be embedded into each other it might be
 possible to induce the quantization of the modified bracket from the 
usual one's.
Nevertheless, it might be worth to see if it is possible to modify the
 constructions used so far for the quantization of Nambu's mechanics.

Nambu proposed a generalization of the Heisenberg commutation relation:
\begin{equation}\label{q1}
cI=\bigl[\hat{A_{1}},\hat{A_{2}},\hat{A_{3}}\bigr]{}=
\sum_{\pi{}\in{}S_{3}}\epsilon(\pi{})
\hat{A}_{\pi(1)}\hat{A}_{\pi(2)}\hat{A}_{\pi(3)}.
\end{equation}
Takhtajan \cite{T} constructed a representation for this relation and its 
generalization.This alternating product might be suitable for Nambu's
 mechanics. However, in the terms of the modified bracket $[f,g,h]$ on
 ${\Bbb{R}}^{2}$
$f,g$ and $h$ play a somewhat different role, so probably the following
 ternary product might be more appropriate:
\begin{equation}\label{E:q2}
\bigl[\hat{A_{1}},\hat{A_{2}},\hat{A_{3}}\bigr]{}
=\rho{}(\hat{A_{1}})\bigl[\hat{A_{2}},\hat{A_{3}}\bigr]{}+
\rho{}(\hat{A_{2}})\bigl[\hat{A_{3}},\hat{A_{1}}\bigr]{}+
\rho{}(\hat{A_{3}})\bigl[\hat{A_{1}},\hat{A_{2}}\bigr]{},
\end{equation}
where $\rho{}$ is some linear functional.

Deformation quantization \cite{BFFS} of an algebra satisfying
 certain identities
 is a somewhat better defined task. For Nambu's mechanics a quite
 novel approach was developed in the papers \cite{DF,DFST}, while the 
straightforward generalization of Weyl's quantization does not
 seems to work very well. In our case, the analog of Weyl's 
quantization could be the following deformed triple product:
\begin{align}
&\bigl({} e,f,g\bigr){}_{h}^{m}=
\pi{}\circ{}\exp{(hD)}\bigl({}e\otimes{}f\otimes{}g\bigr){} \notag{} \\
&D=\partial{}_{x}\wedge{}\partial{}_{y}\otimes{}I+
I\otimes{}\partial{}_{x}\wedge{}\partial{}_{y}+
\bigl({}\partial{}_{x}\otimes{}I\otimes{}\partial{}_{y}-
        \partial{}_{y}\otimes{}I\otimes{}\partial{}_{x} \bigr){},
\end{align}
where $\pi{}$ is the restriction to the diagonal. However, it is
unlikely that this deformed product satisfies the FI.

The n-norm preserving nature of the modified evolution equation deserve some 
attention. Since neither the 1-norm nor the 2-norm is invariant, the usual
probabilistic an quantum mechanical interpretations might be
inappropriate in this case. 
Although the generator $T_{H_{1},\dots{},H_{n}}$  of the evolution
 is a linear operator, in a theory similar to quantum mechanics one can not
 expect to use linear operators if a p-norms $p\neq{}2$ need to be
 preserved. Indeed, the only linear operators leaving the norm 
$\Vert{}x\Vert{}_{p}^{p}= \sum{}|x_{i}|^{p}$ are 
basically the permutations of
 the basis vectors and sign changes.

The modified bracket has an interesting relation to cyclic cocycles
\cite{connes}. On the algebra $C^{\infty{}}(M)$ of smooth
functions of an $n$-dimensional manifold a cyclic cocycle is given by
\begin{equation}
\tau{}(f_{0},\dots{},f_{n})=
\int_{M^{n}} f_{0}df_{1}\wedge{}\dots{}\wedge{}df_{n}.
\end{equation}
This expression is proportional to the integral of the modified Nambu
bracket of $f_{0},\dots{},f_{n}$ over $M^{n}$ for the Nambu $n$-vector
$\eta{}$ satisfying $vol(\eta{})=1$.
\begin{equation}
\tau{}(f_{0},\dots{},f_{n})=
\frac{1}{n}\int_{M^{n}}[f_{0},\dots{},f_{n}] dvol.
\end{equation}

As our last remark, we show that the space of abelian gauge fields 
$A_{i}(x)$
over a three-manifold carries a fairly natural Nambu structure. We
define a trilinear alternating  bracket by
\begin{equation}
{A_{i}(x),A_{j}(y),A_{k}(z)}=
c(x)\delta{}^{3}(x-y)\delta{}^{3}(y-z),
\end{equation}
where $c(x)$ is some nonzero function.This bracket can be extended for
polynomial functions of $A$ by Leibnitz' rule. The FI is satisfied,
since the three-vector generating this bracket is the integral of the
three-vectors
\begin{equation}
\eta{}(x)=c(x)\frac{\delta{}}{\delta{}A_{1}(x)}\wedge{}
\frac{\delta{}}{\delta{}A_{2}(x)}\wedge{}
\frac{\delta{}}{\delta{}A_{1}(x)},
\end{equation}
and $\eta{}(x)$ has the same form as the standard Nambu three-vector
$\partial{}_{1}\wedge{}\partial{}_{2}\wedge\partial{}_{3}$ on
${\Bbb{R}{}}^{3}$. Gauge invariant functions form a closed subalgebra
with respect to this bracket. The variation 
$\delta{}_{\delta(x)}A_{i}(y)=\partial{}_{x_{i}}\delta{}(x-y)$ of
$A$ by a gauge transformation can be expressed as
\begin{equation}
\delta{}_{\delta(x)}A_{i}(y)=
\partial{}_{x_{i}}\delta{}(x-y)=
\frac{1}{4c(x)}\epsilon{}_{lmn}\{F_{lm}(x),A_{m}(x),A_{i}(y)\}.
\end{equation}
If $\Phi{}_{i}[A],\,i=1,2,3$ are gauge invariant functionals, then so
is $\Psi{}[A]=\{\Phi{}_{1},\Phi{}_{2},\Phi{}_{3}\}$, since the FI
implies that
\begin{equation}
\begin{aligned}
\delta{}_{\delta(x)}\Psi{}&=
\frac{1}{4c}\epsilon{}_{lmn}
\{F_{lm}(x),A_{n},\Psi{}_{1}\}\\
&=\frac{1}{4c}\epsilon{}_{lmn}
\biggl({}
\{\{F_{lm}(x),A_{n},\Phi{}_{1}\},\Phi{}_{2},\Phi{}_{3}\}+
\{\Phi{}_{1}\{F_{lm}(x),A_{n},\Phi{}_{2}\},\Phi{}_{3}\}\\
&\phantom{----}+
\{\Phi{}_{1},\Phi{}_{2},\{F_{lm}(x),A_{n},\Phi{}_{3}\}
\biggr){}=0.
\end{aligned}
\end{equation}
Unfortunately,
the natural generalization of this bracket for  nonabelian gauge fields is
not Nambu type:
\begin{equation}
\{A_{i}^{a}(x),A_{j}^{b}(y),A_{k}^{c}(z)\}=
c(x)\epsilon(x){}f^{abc}\delta{}^{3}(x-y)\delta{}^{3}(y-z),
\end{equation}
where $f^{abc}$ is the antisymmetric structure constant of a compact
Lie-group, as this bracket is symmetric in its arguments,

\end{document}